\documentclass[PRB,reprint,amsmath,amssymb,superscriptaddress,longbibliography,floatfix]{revtex4-2}

\usepackage[utf8]{inputenc}  
\usepackage[T1]{fontenc}     
\usepackage{lmodern}         
\DeclareUnicodeCharacter{00F6}{\"o}
\usepackage[T1]{fontenc}
\usepackage{textcomp}

\usepackage{graphicx}        
\usepackage{dcolumn}         
\usepackage{bm}              
\usepackage{booktabs}        
\usepackage{multirow}        
\usepackage{amsmath} 
\usepackage{hyperref}
\hypersetup{
    colorlinks=true,        
    linkcolor=blue,         
    citecolor=blue,         
    urlcolor=blue,          
    breaklinks=true         
}

\usepackage{amssymb}
\usepackage{amsmath}

\usepackage{xcolor}

\begin{document}

\title{Field-induced magnetic phase transitions and transport anomalies in GdAlSi}

\author{Zheng Li}
\affiliation{School of Physics, Zhejiang University, Hangzhou 310058, China}

\author{Sheng Xu}
\affiliation{School of Physics, Zhejiang University, Hangzhou 310058, China}
\affiliation{School of Physics, Zhejiang University of Technology, Hangzhou 310023, China}
\email{shengxu@zjut.edu.cn}

\author{Yi-Yan Wang}
\affiliation{ Anhui Key Laboratory of Magnetic Functional Materials and Devices, Institute of Physical Science and Information Technology,
Anhui University, Hefei 230601, China}

\author{Tian-Hao Li}
\affiliation{School of Physics, Zhejiang University, Hangzhou 310058, China}
\author{Shu-Xiang Li}
\affiliation{School of Physics, Zhejiang University, Hangzhou 310058, China}
\author{Jin-Jin Wang}
\affiliation{School of Physics, Zhejiang University, Hangzhou 310058, China}
\author{Jun-Jian Mi}
\affiliation{School of Physics, Zhejiang University, Hangzhou 310058, China}
\author{Qian Tao}
\affiliation{School of Physics, Zhejiang University, Hangzhou 310058, China}

\author{Zhu-An Xu}
\affiliation{School of Physics, Zhejiang University, Hangzhou 310058, China}
\affiliation{ State Key Laboratory of Silicon and Advanced Semiconductor Materials, Zhejiang University, Hangzhou 310027, China}
\affiliation{Hefei National Laboratory, Hefei 230088, China}
\email{zhuan@zju.edu.cn}

\date{\today}

\begin{abstract}

Magnetic topological materials hosting non-zero Berry curvature have emerged as a focus of intensive research due to their exceptional magnetoelectric coupling phenomena and potential applications in next-generation spintronic devices. In this work, we successfully synthesized high-quality GdAlSi single crystals, a prototypical member of $R$Al$X$ ($R$ = rare earth elements; $X$ = Si/Ge)  family that has been theoretically predicted to sustain a non-trivial Weyl semimetal state. Through systematic investigations of magnetic and transport properties, we identified two successive antiferromagnetic transitions at critical temperatures $T_{N1}$ $\sim$ 31.9 K and $T_{N2}$ $\sim$ 31.1 K, as evidenced by temperature-dependent resistivity, magnetic susceptibility, and specific heat measurements. Notably, applied magnetic fields exceeding 8 T induce a third magnetic transition ($T_{N3}$), generating a cascade of metamagnetic transitions that collectively form a dendritic phase diagram. This complex magnetic behavior is attributed to the interplay between localized Gd-$4f$ moments and itinerant conduction electrons, possibly mediated by Dzyaloshinskii-Moriya interactions. Transport measurements revealed striking stepwise anomalies in magnetoresistance when crossing phase boundaries, accompanied by pronounced hysteresis loops arising from magnetic moment flopping processes. Our results not only establish GdAlSi as a rich platform for investigating correlated topological states, but also demonstrate its potential for engineering topological phase transitions through magnetic symmetry manipulation in Weyl semimetals.
\end{abstract}


\maketitle

\section{ Introduction}
Weyl semimetals (WSMs) are an important class of topological semimetals characterized by the presence of stable Weyl nodes in their band structure. The realization of this phase requires either breaking space-inversion (SI) \cite{lv2015TaAsPRX, 
 yangTaAs_2015NP, xu2015TaAs-Science} or time-reversal (TR) \cite{nakatsuji2015Nature-Mn3SnAHE,Fe3GeTe2_AHE_2018NM, Co2MnGa_ANE_2018NP,Co2MnGa_2019Science,Co3Sn2S2_ARPES2019Science,Co3Sn2S2_AHE_2018NP} symmetry. The WSMs have been well studied in various types of nonmagnetic materials with specific crystalline symmetry since its initial discovery in the TaAs family materials \cite{lv2015TaAsPRX, yangTaAs_2015NP,xu2015TaAs-Science,hu2016SciRepTaP}, and in the ferromagnetic materials like Co$_3$Sn$_2$S$_2$, Co$_2$MnGa \cite{Co3Sn2S2_ARPES2019Science,Co2MnGa_2019Science}. 
However, in collinear
 antiferromagnetic (AFM) materials, the combined $\mathcal{IT}$ ($\mathcal{I}$ represents the spatial inversion, and $\mathcal{T}$  represents time reversal) typically enforces that the Berry curvature vanishes across the entire first Brillouin zone \cite{AFM2022NMreview,AFM2023npjAM}. This symmetry restriction suppresses the observation of topologically non-trivial transport phenomena. 

Recently, the $\mathit{R}\text{Al}\mathit{X}$ (\textit{R}  = rare earth elements; \textit{X} = Si/Ge) family materials,  characterized by noncentrosymmetric crystal structures and intrinsic 4\textit{f}-electron magnetism \cite{chang2018PRB-RAlGe},  has emerged as a paradigm for exploring symmetry-breaking effects in topological matter. In nonmagnetic LaAlSi/Ge systems where SI symmetry is exclusively broken, combined angle-resolved photoemission spectroscopy and quantum oscillation studies have conclusively demonstrated the coexistence of type-I and type-II Weyl fermions \cite{xu2017SciAdv-LaAlGe-ARPES,su2021PRB-LaAlSi-SdH}. Their magnetic counterparts  $\mathit{R}\text{Al}\mathit{X}$ (\textit{R} = Ce, Pr, Nd, Sm, Gd; \textit{X} = Si, Ge) further break TR symmetry through 4\textit{f}-electron ordering, creating an ideal testbed for studying intertwined topology and magnetism. Notably, experimental observations reveal the singular angular magnetoresistance in CeAlGe \cite{suzuki2019Sciense-CeAlGe-SAMR}, the loop-shaped Hall effect in CeAlSi \cite{yang2021PRB-CeAlSi-LHE}, and possible axial gauge fields in PrAlGe \cite{destraz2020npj-PrAlGe-transport}, demonstrating the rich diversity of this material system.
Recent advances highlight how Weyl fermions actively participate in magnetic exchange processes, generating unprecedented spin-texture-topology couplings.
Particularly, helical magnetism in NdAlSi and SmAlSi, arising from Fermi surface nesting and Weyl-induced Dzyaloshinskii–Moriya (DM) interactions, respectively \cite{gaudet2021NM-NdAlSi,wang2022PRB-NdAlSi-phasediagram,yao2023PRX-SmAlSi-THE}, demonstrates that 4\textit{f}-itinerant electron hybridization can drive complex magnetic phase diagrams featuring multiple zero-field transitions and field-induced metamagnetism.
Within this context, GdAlSi stands out as a critical system due to its high-spin Gd$^{3+}$ (S = 7/2) moments and enhanced exchange interactions.

We grew high quality GdAlSi single crystals and performed refined thermodynamic and transport measurements to unravel its complex magnetic landscape. While previous work \cite{laha2024PRB, GAS2024CPB} reported single magnetic transition ($T_{N1}$ $\sim$ 32 K)through susceptibility measurements, our resistivity and high-resolution specific heat data reveal an additional transition at $T_{N2}$ $\sim$ 31.1 K.

Furthermore, multiple field-induced transitions emerge when applying magnetic fields parallel to the \textit{c}-axis, leading to a dendritic magnetic phase diagram. This intricate pattern may potentially stem from the competing interplay among DM interactions, localized 4$f$ electrons, and itinerant electron states.
Our measurements reveal hysteretic phase transitions in the isothermal field dependence of the longitudinal resistivity. In contrast, corresponding features in the Hall resistivity and magnetic susceptibility measurements show markedly weaker magnitudes, suggesting that the magnetic field-induced phase transitions may involve only minor changes in the magnetic structure, occurring within the broader context of AFM ordering.


\begin{figure}[htb]
	\centering
	\includegraphics[width=0.45\textwidth]{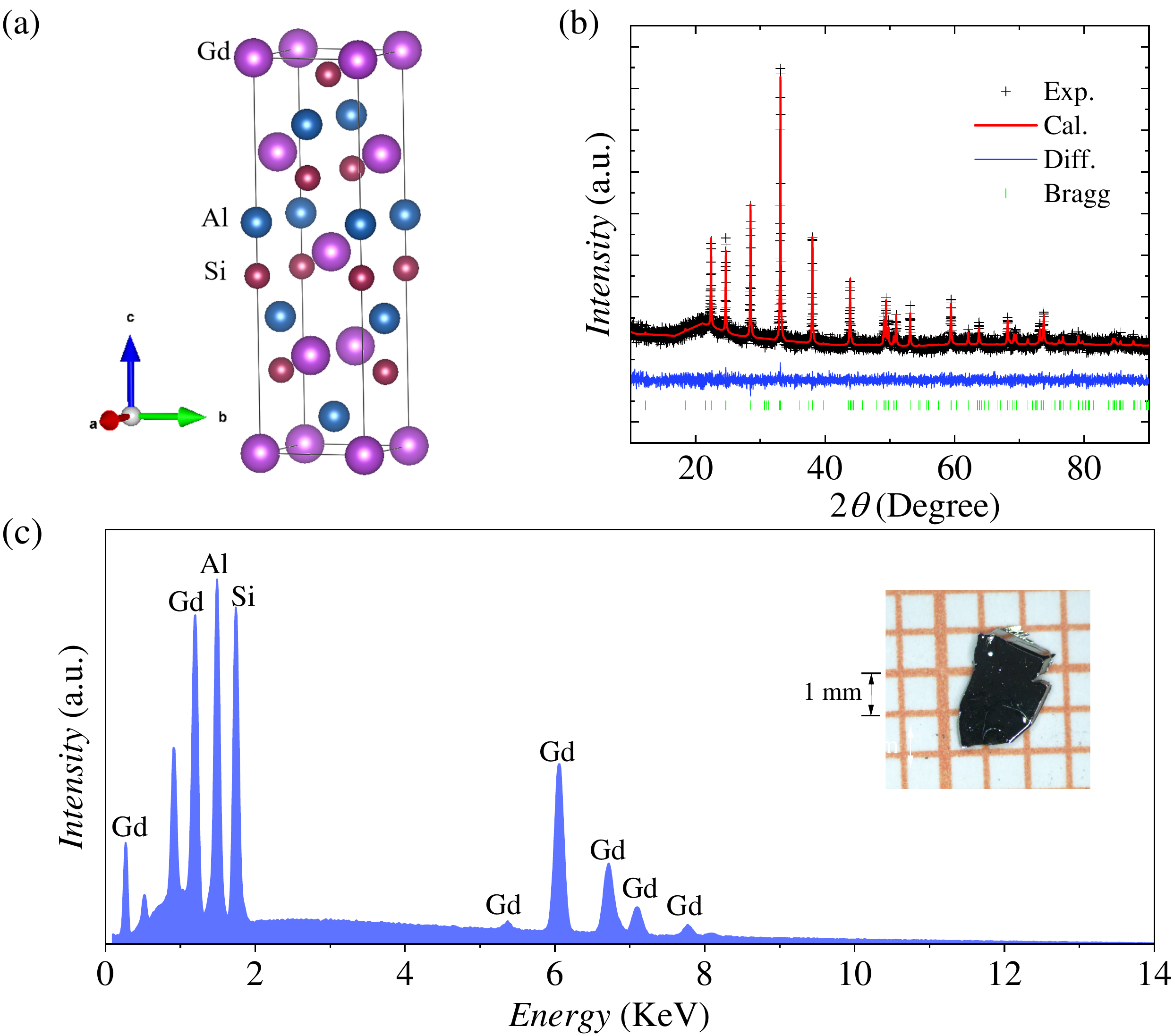}
	\caption{(a) Unit cell of GdAlSi. 
		(b)  Powder XRD pattern measured at room temperature.
		(c)  EDS spectrum of the as-grown single crystal (inset: photograph of the crystal).} 
	\label{fig1}
\end{figure}

\section{Experimental details}
GdAlSi single crystals were grown using a self-flux method. Gadolinium ingots, aluminum pellets, and silicon powder were mixed thoroughly in the ratio of 1:10:1 in an alumina crucible and sealed in an evacuated quartz tube; the tube was heated to 1423 K and maintained at that temperature for 24 hours. Subsequently, it was slowly cooled to 1073 K at a rate of 2 K/h. The excess Al was removed by centrifuging at the end of the growing process. After this process, plate-like single crystals could be harvested. Powder x-ray diffraction (XRD) patterns were collected from the PANalytical x$-$ray diffractometer(Model EMPYREAN) with monochromatic Cu $K\alpha_1$ radiation. The refinement of the data was performed using fullprof software.
The atomic composition of the obtained single crystals was measured by energy-dispersive x-ray spectroscopy (EDS).

The standard six terminal method was applied in the resistivity and Hall effect measurements. All the magnetic and transport measurements were performed on the Quantum Design Magnetic Property Measurement System (QD MPMS$-$5 T) and the Quantum Design Physical Property Measurement Systems (QD PPMS$-$14T) with a vibrating sample magnetometer (VSM) option.

\section{ Results and discussion}

\begin{figure}[htbp]
	\centering
	\includegraphics[width=0.48\textwidth]{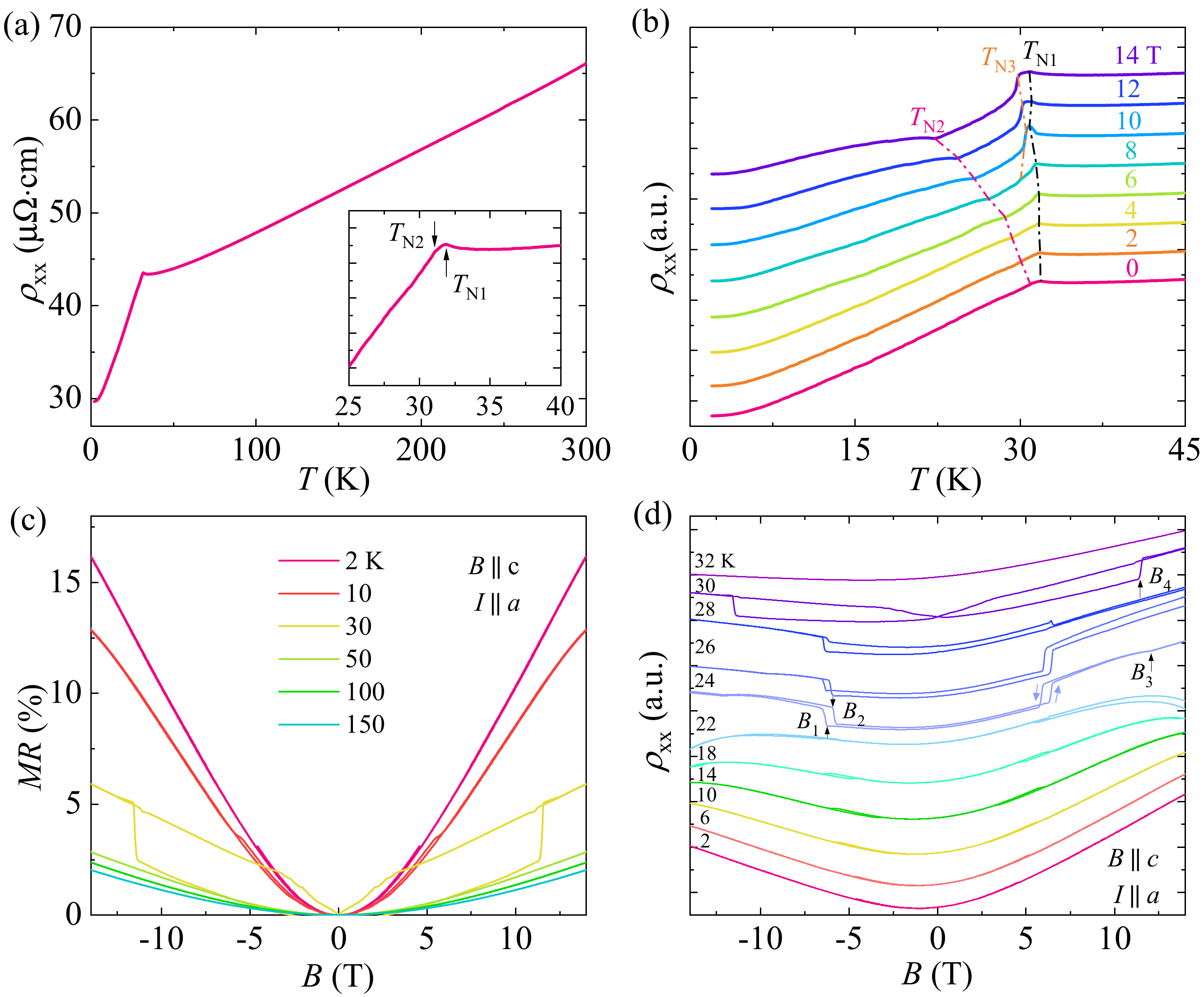}
	\caption{(a) Temperature dependence of $\rho_{xx}$. Inset is the enlargement of the $\rho_{xx}$ around 32 K. (b) Temperatures dependence of $\rho_{xx}$ at different fields. (c) Field dependence of the MR at 
 different temperatures with the $B \parallel c$ configuration. (d) Field dependence of the original data $\rho_{xx}$ under various temperatures between 2 K and 32 K. Arrows mark the critical field positions ($B_1$-$B_4$) corresponding to phase transitions identified from magnetoresistance anomalies.
	}
	\label{fig2}
\end{figure}%

The crystal of GdAlSi exhibits a tetragonal LaPtSi-type crystal structure with noncentrosymmetric space group $I4_1md$, as displayed in Fig.~\ref{fig1}(a). Figure~\ref{fig1}(b) displays the powder XRD patterns, all the diffraction peaks can be well indexed. The refined lattice parameters are $a = b = 4.124 \, \text{\AA}$ and $c = 14.418 \, \text{\AA}$, respectively, consistent with those reported in the literature \cite{laha2024PRB}. The EDS spectra is exhibited in Fig.~\ref{fig1}(c) with composition of about 34.32\%(Gd), 33.14\%(Al), and 32.54\%(Si), respectively, closely to the ratio of Gd: Al: Si = 1: 1: 1.

The temperature-dependent resistivity $\rho_{xx}$ is exhibited in Fig.~\ref{fig2}(a). In addition to the reported AFM phase transition around $T_{N1}$$\sim31.9\,$K \cite{laha2024PRB, GAS2024CPB}, another kink was observed at T$_{N2}$$\sim31.1\,$K. Figure~\ref{fig2}(b) displays the temperature-dependent resistivity under various fields. With increasing magnetic field, $T_{N1}$ remains nearly unchanged, while T$_{N2}$ is suppressed to lower temperatures.
Furthermore, another new transition $T_{N3}$ was induced when the field exceeds 8 T. The field-dependent isothermal magnetoresistances (MR) defined as $MR=100\%\times(\rho{_{xx}}(B)-\rho{_{xx}}(0))/\rho{_{xx}}(0)$ with the $B \parallel c$ configuration are shown in Fig.~\ref{fig2}(c). The MR increases with the increasing field and reaches approximately 16\% at 2~K under magnetic field of 14 T. 

Interestingly, below $T_{N1}$, the MR exhibits a hysteresis loop as shown in Fig.~\ref{fig2}(d). For temperatures below 28~K, the transition fields during increasing and decreasing field sweeps are labeled as $B_1$ and $B_2$, respectively, with the additional anomaly emerging at 24~K assigned to $B_3$. 
In addition to these step-like features at the critical fields, a slight asymmetry is also noticeable in the low-field MR curves at low temperatures (e.g., $T = 26$~K).  Such an asymmetry is often attributed to the history-dependent pinning and motion of magnetic domain walls~\cite{Co_Ybranch2012APL}. Notably, the sequential transitions observed during field sweeps at 30~K exhibit enhanced hysteresis below the transition field $B_4$.
At specific temperatures, the MR in the field-increasing and field-decreasing branches fails to fully recover its initial state at high fields (e.g., 14 T), accompanied by persistent irreversibility. 
This behavior is indicative of a first-order magnetic phase transition (FOMPT) driven by a metamagnetic transition, with the hysteresis of magnetoresistivity below $T_{N1}$ likely originating from magnetic moment flopping during the magnetization process, as observed in pyrochlore systems such as Nd$_2$Ir$_2$O$_7$ \cite{Nd2Ir2O72015PRL}.
Such hysteresis has also been reported in its isostructural counterparts CeAlGe/Si \cite{yang2021PRB-CeAlSi-LHE,lyk2023pressureCeAlGe}.
The evolution of these phase boundaries will be explicitly illustrated in the detailed phase diagram presented later.

\begin{figure}[htb]
	\centering
	\includegraphics[width=0.48\textwidth]{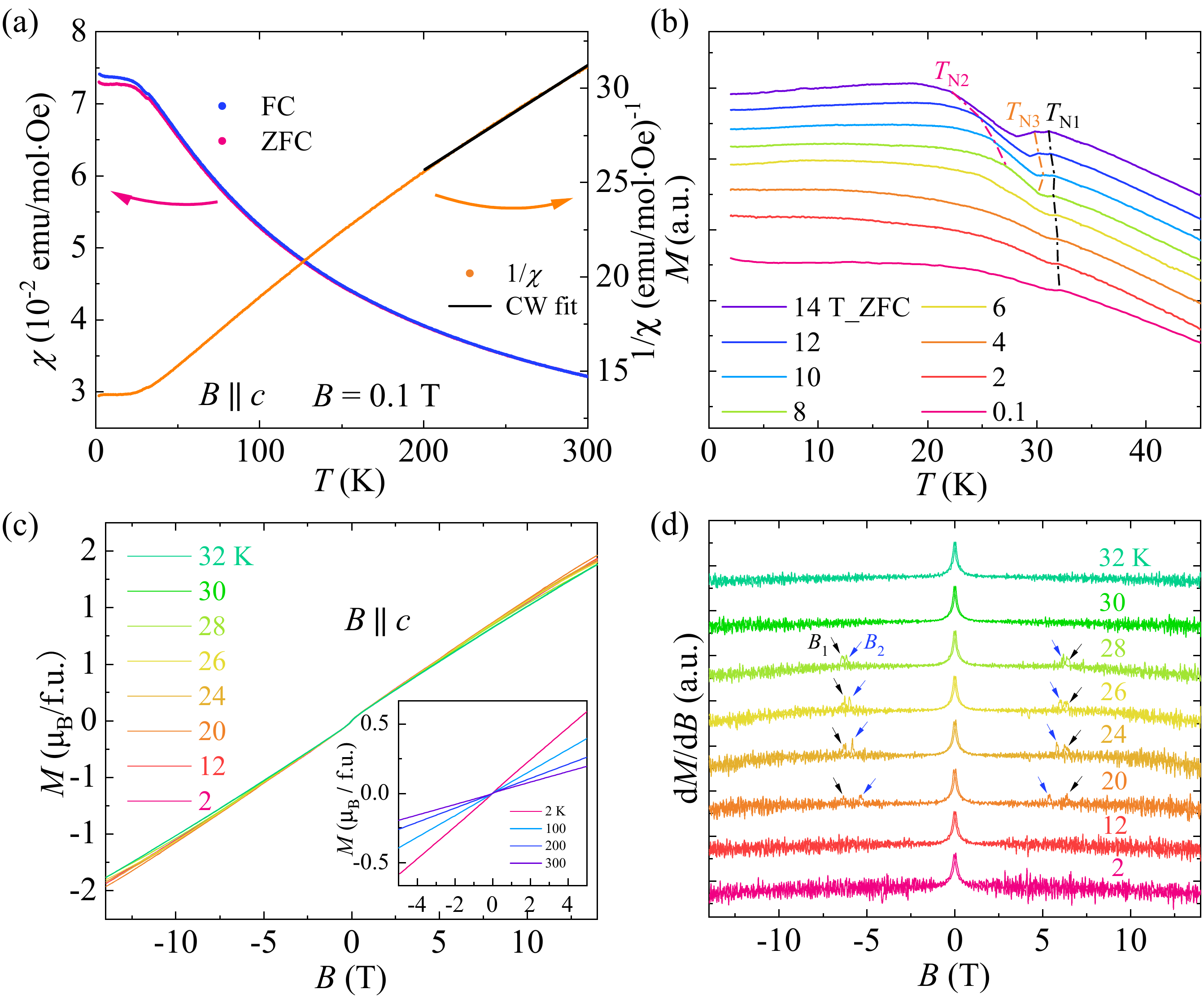}
	\caption{(a) Temperature dependence of magnetic susceptibility $\chi_c$ and 1/$\chi_c$ for $B \parallel c$ measured in an applied field of 0.1~T. 
    (b) Temperature-dependent magnetic susceptibility under different fields. (c) Isothermal magnetization curves measured under applied fields of \textit{B} = -14~T to 14 T at temperatures 2 - 32~K. The inset displays corresponding curves at 2 - 300 K under \textit{B} = ±5~T. (d) Field derivative of magnetization ($dM/dB$) versus magnetic field (\textit{B}) measured at temperatures 2 - 32~K under applied fields up to \textit{B} = ±14~T. Labels $B_1$ and $B_2$ (black/blue arrows) denote anomalies emerging during ascending/descending field sweeps, respectively.
     } 
	\label{fig3}
\end{figure}%

The detailed magnetization measurements were performed to further study the novel magnetic properties in GdAlSi. The temperature dependence of the magnetic susceptibility $\chi$ collected with the $B \parallel c$ configuration ($B=0.1$ T) is shown in Fig.~\ref{fig3}(a). The zero-field-cooled (ZFC) and field-cooled (FC) curves exhibit a slight deviation at low temperatures, which may be attributed to a spin-glass state~\cite{destraz2020npj-PrAlGe-transport,lyu2020PRBPrAlSi} or domain rotation during the cooling process \cite{RCo5Ga7-2004DWZFCFC}.

Notably, the magnetic behavior below $T_\text{N1}$ suggests a complex AFM ground state. 
This is supported by the weak magnetic anisotropy~\cite{laha2024PRB, GAS2024CPB} and a subtle upturn in the susceptibility upon cooling (Fig.~\ref{fig3}(a)), which deviates from the behavior of a simple collinear AFM~\cite{HoSbAFM}. 
This picture is consistent with the behavior of the isostructural spiral magnet SmAlSi~\cite{yao2023PRX-SmAlSi-THE} and is compellingly confirmed by a recent resonant elastic X-ray scattering (REXS) study that identified a cycloidal magnetic structure in GdAlSi~\cite{GdAlSiREXS2025arxiv}. 
At high temperature, 1/$\chi$ follows the Curie-Weiss law, 
\begin{equation}
\chi = \frac{C}{T - T_{\theta}}
\end{equation}
where $T_{\theta}$ is the critical temperature and $C$ is the Curie-Weiss constant. From the fitting in the temperature range of 200 K to 300 K, the effective moment is found to be $7.68\,\mu_{B}/Gd$, which is close to the effective magnetic moment of Gd$^{3+}$.  Figure~\ref{fig3}(b) presents the temperature-dependent magnetization at various fields for the $B \parallel c$ configuration. Consistent with the observations of the resistivity measurements shown in Fig.~\ref{fig1}(c), the $T_{N1}$ remains almost unchanged, while $T_{N2}$ shifts to a lower temperature as the magnetic field increases. The third field-induced transition, characterized by the critical temperature $T_{N3}$ was observed above 8 T, and is clearly highlighted by an additional peak, as shown in Fig.~\ref{fig3}(b).

To investigate the metamagnetic transition observed in the MR, systematic field-dependent magnetization measurements were performed, with a focus on temperatures below $T_{N1}$. Figure~\ref{fig3}(c) demonstrates a characteristic quasi-linear field dependence of magnetization without saturation behavior up to 14 T.
Below $T_{N1}$, the magnetization shows little variation with temperature, reaching approximately 1.8~$\mu_{B}$/f.u. at 2~K and 14~T. It is significantly lower than the expected moment of Gd$^{3+}$ free-ion moment ($\sim 7.94\mu_{B}$). 
While the magnetization curve in Fig.~\ref{fig3}(c)  exhibits minimal variation over the field range, the first-derivative analysis ($dM/dB$) in Fig.~\ref{fig3}(d) reveals two distinct peaks at $B_1$ and $B_2$, observed below 28~K. These critical fields exhibit precise correspondence with the step-like anomalies previously identified in magnetoresistance measurements, though the transition amplitudes are significantly attenuated in the magnetization data.
Detailed analysis of the corresponding Hall signatures and comparative transport-magnetic correlations will be addressed hereafter. 

In stark contrast, the MR at 30~K exhibits significantly enhanced hysteresis  below $B_4$, as shown in Fig.~\ref{fig2}(d).
However, this phenomenon shows no corresponding signatures in magnetic susceptibility data in Figs.~\ref{fig3}(c)(d).
Similar field-induced resistive anomalies decoupled from bulk magnetism were reported in  heavy fermion systems \cite{UIrSi32019magnetotransport, CePtSn2010MR}. 
These results suggest that the metamagnetic transitions are relatively moderate, occurring within the broader framework of AFM ordering.
This scenario is highly consistent with the cycloidal ground state discussed previously~\cite{GdAlSiREXS2025arxiv}, 
as such a complex spin texture can be distorted by an applied field without causing a large change in the net magnetization. 
In addition, the sharp peak centered near $B = 0$~T in the $\mathrm{d}M/\mathrm{d}B$ data below $T_{N1}$ (Fig.~\ref{fig3}(d)) reflects the initial response of this non-collinear state, possibly originating from spin canting or the alignment of metastable magnetic domains~\cite{nakatsuji2015Nature-Mn3SnAHE, GdAlSiREXS2025arxiv}.

\begin{figure}[h]
	\centering
	\includegraphics[width=0.48\textwidth]{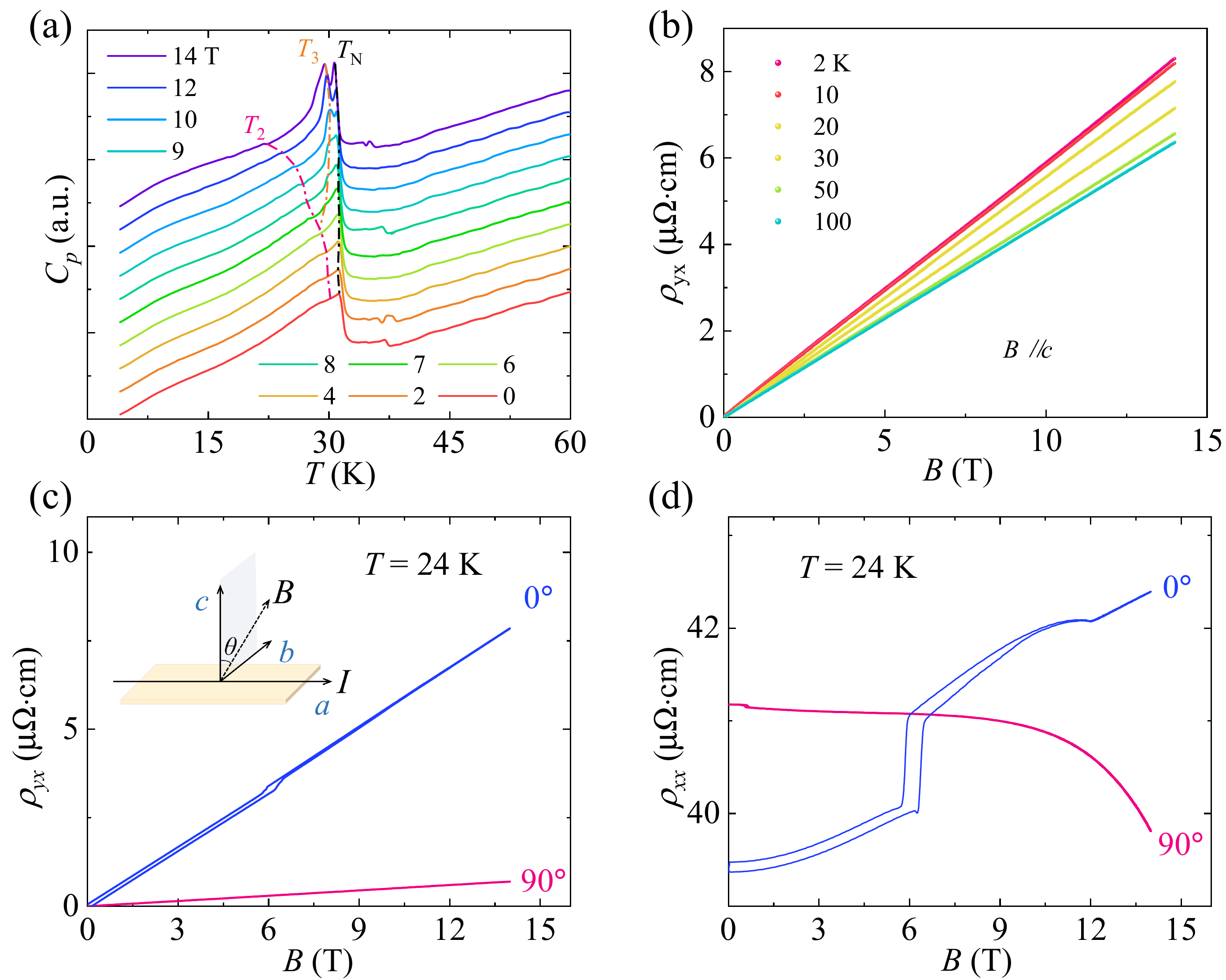}
	\caption{(a) Temperature dependence of the  specific heat $C(T)$ under different fields. (b) Magnetic field dependence of Hall resistivity at various temperatures. (c) Hall resistivity ($\rho_{yx}$) and (d) magnetoresistivity ($\rho_{xx}$) for magnetic fields applied along the $c$-axis ($\theta = 0^\circ$) and in the $ab$-plane ($\theta = 90^\circ$).
	}
	\label{fig4}
\end{figure}%

\begin{figure}[htb]
	\centering
	\includegraphics[width=0.48\textwidth]{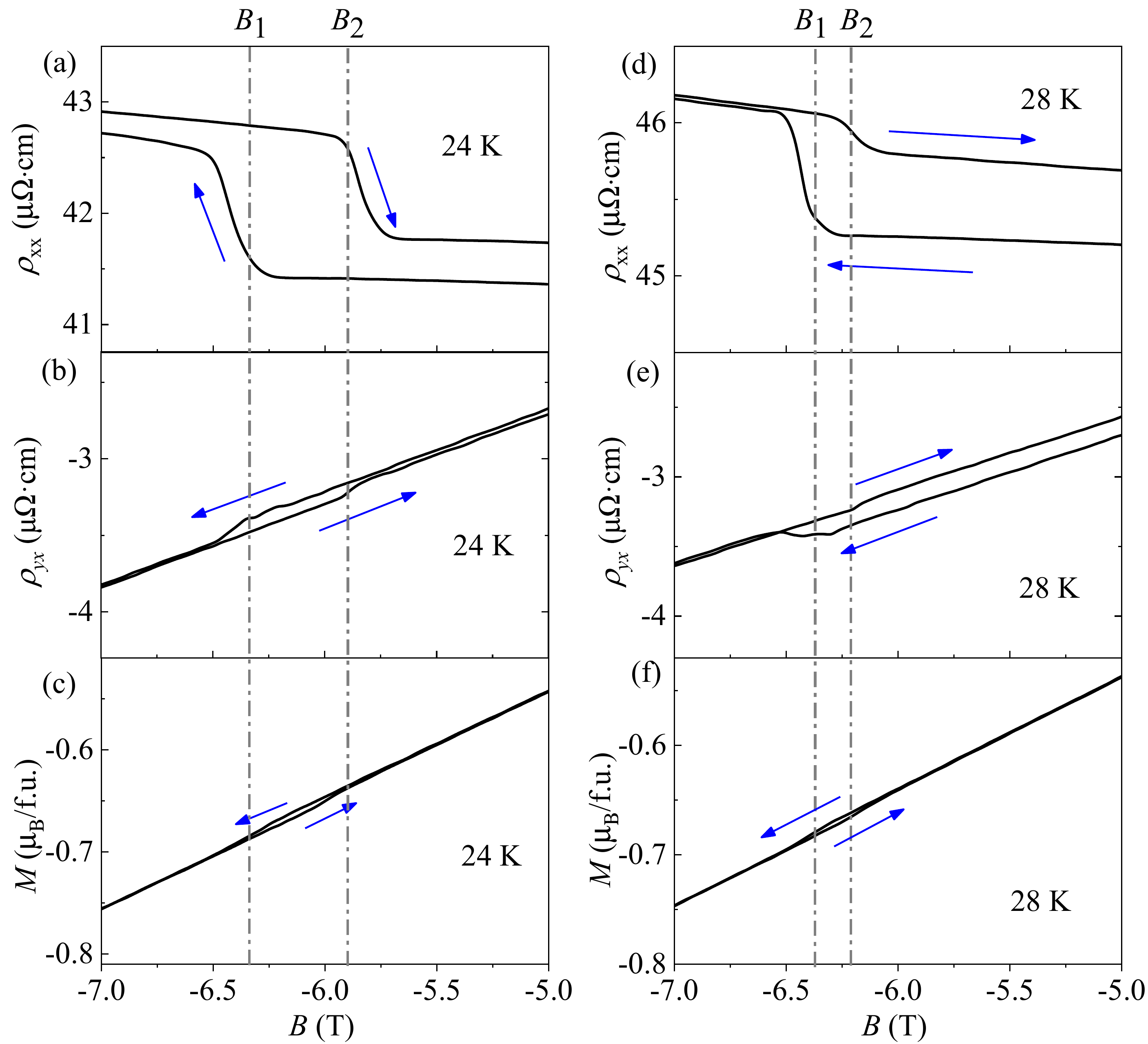}
	\caption{(a)(b)(c), (d)(e)(f) Enlarged plots of $\rho_{xx}$, $\rho_{yx}$,  and $M$ around the critical field $B_1$ and $B_2$ of the metamagnetic transition at 24 K and 28 K, respectively. The blue arrows indicate the direction of the magnetic field sweep.
	}
	\label{fig5}
\end{figure}%

The temperature-dependent specific heat capacity is presented in Fig.~\ref{fig4}(a). As shown, in addition to the AFM transition $T_{N1}$$\sim$31.9 K, there is another shoulder around $T_{N2}$$\sim$31.1 K.
The emergence of field-induced sharp peaks in the specific heat manifests bulk properties, indicating potential transitions among distinct magnetic configurations.
As the magnetic field increases, \( T_{N2} \) shifts to lower temperatures, and a third phase transition, \( T_{N3} \), occurs at approximately 8 T, becoming particularly pronounced at 14 T. These characteristics are consistent with the observation in the MR and magnetization measurements, and indicating that the observed transitions are bulk properties.

The anisotropic nature of the transport phenomena was further examined at $T = 24$~K, as shown in Figs.~\ref{fig4}(c) and~\ref{fig4}(d). 
A stark contrast is observed between the responses for magnetic fields applied along the $c$-axis ($\theta = 0^\circ$) 
and within the $ab$-plane ($\theta = 90^\circ$). 
The prominent hysteretic and step-like features, which are hallmarks of the transport for $B \parallel c$, 
are substantially suppressed and become indiscernible when the field is aligned with the $ab$-plane.

Between 20~K and 30~K, the $\rho_{yx}$ anomaly induced by the field-induced metamagnetic transitions at the critical fields $B_1$ and $B_2$ were also observed. Figures~\ref{fig5}(a)(b)(c) and ~\ref{fig5}(d)(e)(f) present enlarged plots of $\rho_{xx}$, $\rho_{yx}$, and $M$ around the critical field $B_1$ and $B_2$ in 24~K and 28~K, respectively. The anomaly observed in the $\rho_{xx}$ and $\rho_{yx}$ is closely related to the small kinks in magnetization, suggesting that these transport anomalies may be linked to magnetic moment flopping. 

\begin{figure}[htb]
	\centering
	\includegraphics[width=0.48\textwidth]{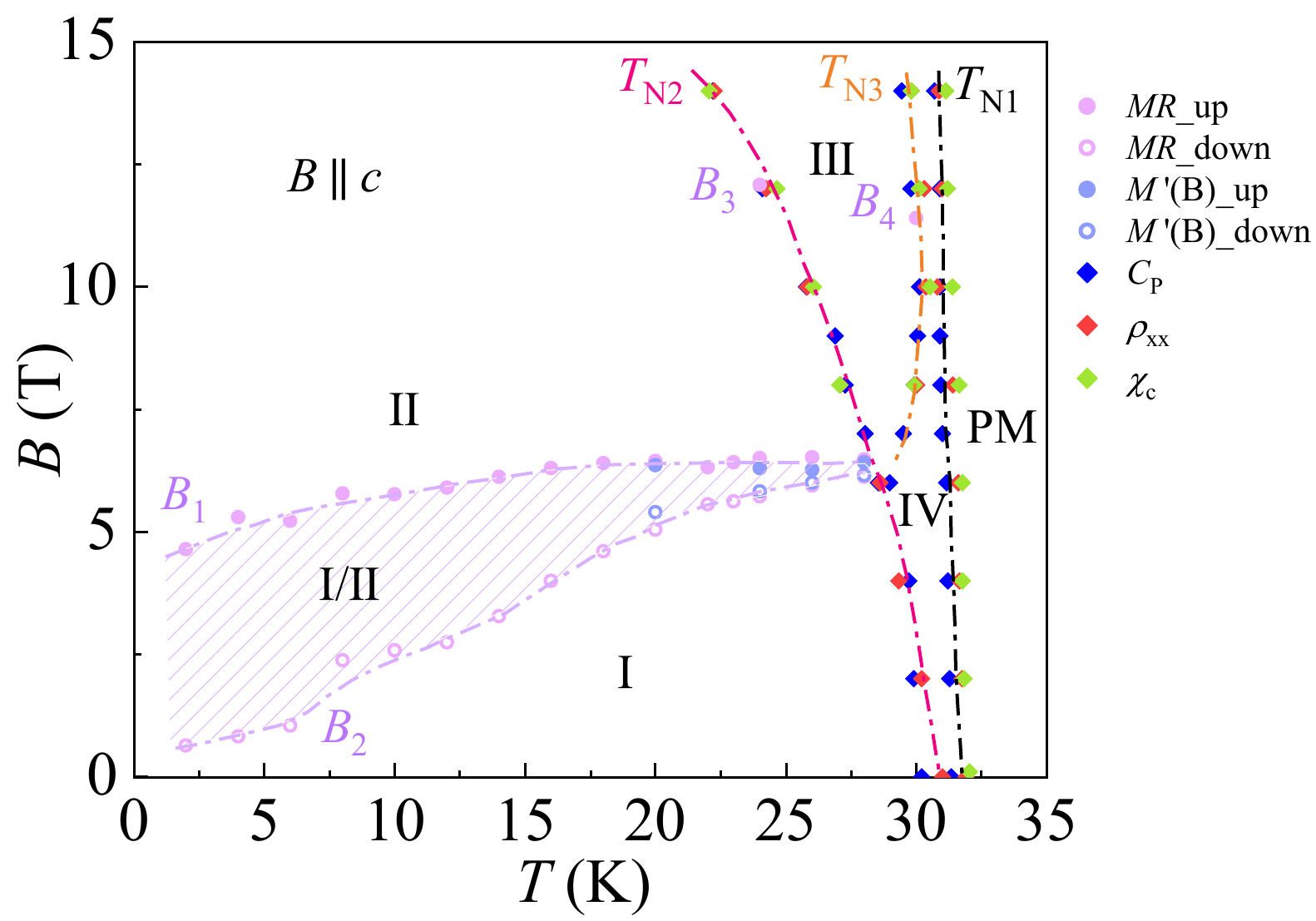}
	\caption{Temperature-magnetic field phase diagram of GdAlSi with the applied field parallel to $c$ axis.}
	\label{fig6}
\end{figure}%

Building on the aforementioned analysis, a $B-T$ phase diagram exhibiting a dendritic structure is constructed. As shown in Fig.~\ref{fig6}, the phase I is bound horizontally by the hysteresis loops in the MR (Fig.~\ref{fig2}(d)) and $dM/dH$ (Fig.~\ref{fig3}(d)).
The solid purple circles between phase I and phase II denote the transition field $B_1$ during the increasing field, while the hollow purple circles indicate the transition field $B_2$ during the decreasing field.
The phase IV is bound vertically by the $T_{N1}$ and $T_{N2}$ in the $\rho_{xx}(T)$, $\chi_{c}(T)$ and $C_{p}(T)$ as shown in Fig.~\ref{fig2}(b), Fig.~\ref{fig3}(b) and Fig.~\ref{fig4}(a).
A field-induced phase III appears above about 6~T located between phase II and phase IV. The field-induced transition $T_{N3}$(orange dotted line) appears to terminate on intersecting $T_{N2}$ around 28 K, which, together with $T_{N2}$, forms the boundary of the field-induced phase III.

We can infer that Phase I represents the initial, complex non-collinear AFM ground state, which is consistent with the cycloidal structure \cite{GdAlSiREXS2025arxiv}. This state is then followed by a metamagnetic transition that leads to the field-induced Phase II.
The region between $B_1$ and $B_2$ defines a hysteretic regime where the dominant phase (I or II) depends on the field sweep direction. Phase II stabilizes above $B_1$ during the increasing field, while Phase I persists below $B_2$ until the decreasing field reverses the transition. 
The phase transition from Phase II to Phase III across the phase boundary determined by $T_{N2}$ is marked by a distinct kink in the MR at 24~K. This anomaly, identified as the critical field $B_3$, is marked in the MR data of Fig.~\ref{fig2}(d).
Crucially, the critical fields $B_4$ extracted from the 30~K magnetoresistance hysteresis coincide with the $T_{N3}$-defined phase boundary, demonstrating that the hysteresis originates from the FOMPT between two competing magnetic structures \cite{Dy3Co2000metamagnetic,UIrSi32019magnetotransport,CePtSn2003neutron,CePtSn2010MR}.

\section{Conclusion}
In summary, we have successfully synthesized high-quality single crystals of GdAlSi, a recently identified compound of the $\mathit{R}\text{Al}\mathit{X}$ family, which is considered to host a magnetic Weyl state. The rich magnetism and electric transport properties, probably influenced by the DM interaction, were systematically investigated. Even under zero magnetic field, two magnetic phase transitions were found, with critical temperatures $T_{N1}$ $\sim$ 31.9~K, and $T_{N2}$ $\sim$ 31.1~K. Moreover, via combined analysis of transport, magnetization, and specific heat measurements, we identified multiple field-induced magnetic phase transitions. These transitions are marked by distinct transport anomalies, including step-like features and hysteresis in magnetoresistance, Hall resistivity, and magnetization, suggesting a coupling between magnetic order reconstruction and spin-dependent electronic transport. Our work demonstrates the sensitivity of electrical transport phenomena in probing magnetic structural transformations and highlight the potential of GdAlSi as a promising candidate for exploring novel magnetic and transport phenomena. Furthermore, GdAlSi may offer a new platform for investigating exotic topological states, warranting further investigation. 




\section*{Acknowledgments}
This work was supported by the National Natural Science Foundation of China (Grant No. 12174334; 12204410), the National Key R\&D Program of China (Grant No.2019YFA0308602), the Innovation program for Quantum Science and Technology(Grant No. 2021ZD0302500), the Zhejiang Provincial Natural Science Foundation of China(Grant No. LMS25A040002) 

\begin{thebibliography}{99}

\bibitem{lv2015TaAsPRX} 
B. Q. Lv, H. M. Weng, B. B. Fu, X. P. Wang, H. Miao, J. Ma, P. Richard, X. C. Huang, L. X. Zhao, G. F. Chen, X. Dai, T. Qian, and H. Ding, 
Experimental discovery of Weyl semimetal TaAs, 
\href{https://doi.org/10.1103/PhysRevX.5.031013}{Phys. Rev. X \textbf{5}, 031013 (2015).}

\bibitem{yangTaAs_2015NP} 
L. X. Yang, Z. K. Liu, Y. Sun, H. Peng, H. F. Yang, T. Zhang, B. Zhou, Y. Zhang, Y. F. Guo, M. Rahn, P. D. Prabhakaran, Z. Hussain, S.-K. Mo, C. Felser, B. Yan, and Y. L. Chen, 
Weyl semimetal phase in the non-centrosymmetric compound TaAs, 
\href{https://doi.org/10.1038/NPHYS3425}{Nat. Phys. \textbf{11}, 728 (2015).}

\bibitem{xu2015TaAs-Science} 
S. Y. Xu, I. Belopolski, N. Alidoust, M. Neupane, G. Bian, C. Zhang, R. Sankar, G. Chang, Z. Yuan, C. Lee, S. Huang, H. Zheng, J. Ma, D. S. Sanchez, B. K. Wang, A. Bansil, F. Chou, P. P. Shibayev, H. Lin, S. Jia, and M. Z. Hasan, 
Discovery of a Weyl fermion semimetal and topological Fermi arcs, 
\href{https://doi.org/10.1126/science.aaa9297}{Science \textbf{349}, 613 (2015).}

\bibitem{nakatsuji2015Nature-Mn3SnAHE} 
S. Nakatsuji, N. Kiyohara, and T. Higo, 
Large anomalous Hall effect in a non-collinear antiferromagnet at room temperature, 
\href{https://doi.org/10.1038/nature15723}{Nature \textbf{527}, 212 (2015).}

\bibitem{Fe3GeTe2_AHE_2018NM} 
K. Kim, J. Seo, E. Lee, K. T. Ko, B. S. Kim, B. G. Jang, J. M. Ok, J. Lee, Y. J. Jo, W. Kang, J. H. Shim, C. Kim, H. W. Yeom, B. Min, B. J. Yang, and J. S. Kim, 
Large anomalous hall current induced by topological nodal lines in a ferromagnetic van der Waals semimetal, 
\href{https://doi.org/10.1038/s41563-018-0132-3}{Nat. Mater. \textbf{17}, 794 (2018).}

\bibitem{Co2MnGa_ANE_2018NP} 
A. Sakai, Y. P. Mizuta, A. A. Nugroho, R. Sihombing, T. Koretsune, M. T. Suzuki, N. Takemori, R. Ishii, D. Nishio-Hamane, R. Arita, P. Goswami, and S. Nakatsuji, 
Giant anomalous Nernst effect and quantum-critical scaling in a ferromagnetic semimetal, 
\href{https://doi.org/10.1038/s41567-018-0225-6}{Nat. Phys. \textbf{14}, 1119 (2018).}

\bibitem{Co2MnGa_2019Science} 
I. Belopolski, K. Manna, D. S. Sanchez, G. Chang, B. Ernst, J. Yin, S. S. Zhang, T. Cochran, N. Shumiya, H. Zheng, B. Singh, G. Bian, D. Multer, M. Litskevich, X. Zhou, S. M. Huang, T. R. Wang, B. Chang, S. Y. Xu, A. Bansil, C. Felser, H. Lin, and M. Z. Hasan, 
Discovery of topological Weyl fermion lines and drumhead surface states in a room temperature magnet, 
\href{https://doi.org/10.1126/science.aav2327}{Science \textbf{365}, 1278 (2019).}

\bibitem{Co3Sn2S2_ARPES2019Science} 
D. F. Liu, A. J. Liang, E. K. Liu, Q. N. Xu, Y. W. Li, C. Chen, D. Pei, W. J. Shi, S. K. Mo, P. Dudin, T. Kim, C. Cacho, G. Li, Y. Sun, L. X. Yang, Z. K. Liu, S. S. P. Parkin, C. Felser, and Y. L. Chen, 
Magnetic Weyl semimetal phase in a Kagom{\'e} crystal, 
\href{https://doi.org/10.1126/science.aav2873}{Science \textbf{365}, 1282 (2019).}

\bibitem{Co3Sn2S2_AHE_2018NP} 
E. Liu, Y. Sun, N. Kumar, L. Muechler, A. Sun, L. Jiao, S. Y. Yang, D. Liu, A. Liang, Q. Xu, J. Kroder, V. S{\"u}{\ss}, H. Borrmann, C. Shekhar, Z. Wang, C. Xi, W. Wang, W. Schnelle, S. Wirth, Y. Chen, S. T. B. Goennenwein, and C. Felser, 
Giant anomalous Hall effect in a ferromagnetic kagome-lattice semimetal, 
\href{https://doi.org/10.1038/s41567-018-0234-5}{Nat. Phys. \textbf{14}, 1125 (2018).}

\bibitem{hu2016SciRepTaP} 
J. Hu, J. Y. Liu, D. Graf, S. M. A. Radmanesh, D. J. Adams, A. Chuang, Y. Wang, I. Chiorescu, J. Wei, L. Spinu, and Z. Q. Miao, 
$\pi$ Berry phase and Zeeman splitting of Weyl semimetal TaP, 
\href{https://doi.org/10.1038/srep18674}{Sci. Rep. \textbf{6}, 18674 (2016).}

\bibitem{AFM2022NMreview} 
L. {\v{S}}mejkal, A. H. MacDonald, J. Sinova, S. Nakatsuji, and T. Jungwirth, 
Anomalous hall antiferromagnets, 
\href{https://doi.org/10.1038/s41578-022-00430-3}{Nat. Rev. Mater. \textbf{7}, 482 (2022).}

\bibitem{AFM2023npjAM} 
P. J. Guo, Z. X. Liu, and Z. Y. Lu, 
Quantum anomalous hall effect in collinear antiferromagnetism, 
\href{https://doi.org/10.1038/s41524-023-01025-4}{npj Comput. Mater. \textbf{9}, 70 (2023).}

\bibitem{chang2018PRB-RAlGe} 
G. Chang, B. Singh, S. Y. Xu, G. Bian, S. M. Huang, C. H. Hsu, I. Belopolski, N. Alidoust, D. S. Sanchez, H. Zheng, H. Lu, X. Zhang, Y. Bian, T. R. Chang, H. T. Jeng, A. Bansil, H. Hsu, J. S. Jia, T. Neupert, H. Lin, and M. Z. Hasan, 
Magnetic and noncentrosymmetric Weyl fermion semimetals in the RAlGe family of compounds (R= rare earth), 
\href{https://doi.org/10.1103/PhysRevB.97.041104}{Phys. Rev. B \textbf{97}, 041104 (2018).}

\bibitem{xu2017SciAdv-LaAlGe-ARPES} 
S. Y. Xu, N. Alidoust, G. Chang, H. Lu, B. Singh, I. Belopolski, D. S. Sanchez, X. Zhang, G. Bian, H. Zheng, M. A. Husanu, Y. Bian, S. M. Huang, C. H. Hsu, T. R. Chang, H. T. Jeng, A. Bansil, T. Neupert, V. N. Strocov, H. Lin, S. Jia, and M. Z. Hasan, 
Discovery of Lorentz-violating type II Weyl fermions in LaAlGe, 
\href{https://doi.org/10.1126/sciadv.1603266}{Sci. Adv. \textbf{3}, e1603266 (2017).}

\bibitem{su2021PRB-LaAlSi-SdH} 
H. Su, X. Shi, J. Yuan, Y. Wan, E. Cheng, C. Xi, L. Pi, X. Wang, Z. Zou, N. Yu, W. Zhan, S. Li, and Y. Guo, 
Multiple Weyl fermions in the noncentrosymmetric semimetal LaAlSi, 
\href{https://doi.org/10.1103/PhysRevB.103.165128}{Phys. Rev. B \textbf{103}, 165128 (2021).}

\bibitem{suzuki2019Sciense-CeAlGe-SAMR} 
T. Suzuki, L. Savary, J. P. Liu, J. W. Lynn, L. Balents, and J. G. Checkelsky, 
Singular angular magnetoresistance in a magnetic nodal semimetal, 
\href{https://doi.org/10.1126/science.aat0348}{Science \textbf{365}, 377 (2019).}

\bibitem{yang2021PRB-CeAlSi-LHE} 
H. Y. Yang, B. Singh, J. Gaudet, B. Lu, C. Y. Huang, W. C. Chiu, S. M. Huang, B. Wang, F. Bahrami, B. Xu, J. Franklin, I. Sochnikov, D. E. Graf, G. Xu, Y. Zhao, C. M. Hoffman, H. Lin, D. H. torchinsky, C. L. Broholm, A. Bansil, and F. Tafti, 
Noncollinear ferromagnetic Weyl semimetal with anisotropic anomalous Hall effect, 
\href{https://doi.org/10.1103/PhysRevB.103.115143}{Phys. Rev. B \textbf{103}, 115143 (2021).}

\bibitem{destraz2020npj-PrAlGe-transport} 
D. Destraz, L. Das, S. S. Tsirkin, Y. Xu, T. Neupert, J. Chang, A. Schilling, A. G. Grushin, J. Kohlbrecher, L. Keller, P. Puphal, E. Pomjakushina, and J. S. White, 
Magnetism and anomalous transport in the Weyl semimetal PrAlGe: possible route to axial gauge fields, 
\href{https://doi.org/10.1038/s41535-019-0207-7}{npj Quantum Mater. \textbf{5}, 5 (2020).}

\bibitem{gaudet2021NM-NdAlSi} 
J. Gaudet, H. Y. Yang, S. Baidya, B. Lu, G. Xu, Y. Zhao, J. A. Rodriguez-Rivera, C. M. Hoffmann, D. E. Graf, D. H. Torchinsky, P. Nikoli{\'c}, D. Vanderbilt, F. Tafti, and C. L. Broholm, 
Weyl-mediated helical magnetism in NdAlSi, 
\href{https://doi.org/10.1038/s41563-021-01062-8}{Nat. Mater. \textbf{20}, 1650 (2021).}

\bibitem{wang2022PRB-NdAlSi-phasediagram} 
J. F. Wang, Q. X. Dong, Z. P. Guo, M. Lv, Y. F. Huang, J. S. Xiang, Z. A. Ren, Z. J. Wang, P. J. Sun, G. Li, and G. F. Chen, 
NdAlSi: A magnetic Weyl semimetal candidate with rich magnetic phases and atypical transport properties, 
\href{https://doi.org/10.1103/PhysRevB.105.144435}{Phys. Rev. B \textbf{105}, 144435 (2022).}

\bibitem{yao2023PRX-SmAlSi-THE} 
X. Yao, J. Gaudet, R. Verma, D. E. Graf, H. Y. Yang, F. Bahrami, R. Zhang, A. A. Aczel, S. Subedi, D. H. Torchinsky, J. Sun, A. Bansil, S. M. Huang, B. Singh, P. Blaha, P. Nikoli{\'c}, and F. Tafti, 
Large topological hall effect and spiral magnetic order in the Weyl semimetal SmAlSi, 
\href{https://doi.org/10.1103/PhysRevX.13.011035}{Phys. Rev. X \textbf{13}, 011035 (2023).}

\bibitem{laha2024PRB} 
A. Laha, A. K. Kundu, N. Aryal, E. S. Bozin, J. Yao, S. Paone, A. Rajapitamahuni, E. Vescovo, T. Valla, M. Abeykoon, R. Jing, W. Yin, A. N. Pasupathy, M. Liu, and Q. Li, 
Electronic structure and magnetic and transport properties of antiferromagnetic Weyl semimetal GdAlSi, 
\href{https://doi.org/10.1103/PhysRevB.109.035120}{Phys. Rev. B \textbf{109}, 035120 (2024).}

\bibitem{GAS2024CPB} 
J. Gong, H. Wang, X. Ma, X. Zeng, J. Lin, K. Han, Y. Wang, and T. Xia, 
Magnetic and electrical transport properties in GdAlSi and SmAlGe, 
\href{https://doi.org/10.1088/1674-1056/ad41ba}{Chin. Phys. B \textbf{33}, 077302 (2024).}

\bibitem{Co_Ybranch2012APL} 
Z. Chen, T. Y. Lin, X. Wei, M. Matsunaga, T. Doi, Y. Ochiai, N. Aoki, and J. P. Bird, 
The magnetic Y-branch nanojunction: Domain-wall structure and magneto-resistance, 
\href{https://doi.org/10.1063/1.4750240}{Appl. Phys. Lett. \textbf{101}, 102403 (2012).}

\bibitem{Nd2Ir2O72015PRL} 
K. Ueda, J. Fujioka, B. J. Yang, J. Shiogai, A. Tsukazaki, S. Nakamura, S. Awaji, N. Nagaosa, and Y. Tokura, 
Magnetic field-induced insulator-semimetal transition in a pyrochlore Nd$_2$Ir$_2$O$_7$, 
\href{https://doi.org/10.1103/PhysRevLett.115.056402}{Phys. Rev. Lett. \textbf{115}, 056402 (2015).}

\bibitem{lyk2023pressureCeAlGe} 
X. He, Y. Li, H. Zeng, Z. Zhu, S. Tan, Y. Zhang, C. Cao, and Y. Luo, 
Pressure-tuning domain-wall chirality in noncentrosymmetric magnetic Weyl semimetal CeAlGe, 
\href{https://doi.org/10.1007/s11433-022-2051-4}{Sci. China Phys., Mech. \& Astron. \textbf{66}, 237011 (2023).}

\bibitem{lyu2020PRBPrAlSi} 
M. Lyu, J. Xiang, Z. Mi, H. Zhao, Z. Wang, E. Liu, G. Chen, Z. Ren, G. Li, and P. Sun, 
Nonsaturating magnetoresistance, anomalous Hall effect, and magnetic quantum oscillations in the ferromagnetic semimetal PrAlSi, 
\href{https://doi.org/10.1103/PhysRevB.102.085143}{Phys. Rev. B \textbf{102}, 085143 (2020).}

\bibitem{RCo5Ga7-2004DWZFCFC} 
H. Chang, Y. Q. Guo, J. K. Liang, and G. H. Rao, 
Magnetic ordering and irreversible magnetization between ZFC and FC states in RCo$_5$Ga$_7$ compounds, 
\href{https://doi.org/10.1016/j.jmmm.2003.12.1316}{J. Magn. Magn. Mater. \textbf{278}, 306 (2004).}

\bibitem{HoSbAFM} 
Z. Xia, H. Chen, Y. Chen, F. Tang, R. Wang, C. Hu, L. Jiang, Y. Fang, Z. Han, and J. Hu, 
Giant anisotropic magnetocaloric effect in antiferromagnetic topological semimetal HoSb, 
\href{https://doi.org/10.1016/j.jallcom.2024.177495}{J. Alloys Compd. \textbf{1010}, 177495 (2025).}

\bibitem{GdAlSiREXS2025arxiv} 
R. Nakano, R. Yamada, J. Bouaziz, M. Colling, M. Gen, K. Shoriki, Y. Okamura, A. Kikkawa, H. Ohsumi, Y. Tanaka, H. Sagayama, H. Nakao, Y. Taguchi, Y. Takahashi, M. Tokunaga, T.-H. Arima, Y. Tokura, R. Arita, J. Masell, S. Hayami, and M. Hirschberger, 
Perfectly harmonic spin cycloid and multi-Q textures in the Weyl semimetal GdAlSi, 
\href{https://arxiv.org/abs/2503.14768}{arXiv:2503.14768 (2025).}

\bibitem{UIrSi32019magnetotransport} 
F. Honda, J. Valenta, J. Prokle{\v{s}}ka, J. Posp{\'i}{\v{s}}il, P. Proschek, J. Prchal, and V. Sechovsk{\'y}, 
Magnetotransport as a probe of phase transformations in metallic antiferromagnets: The case of UIrSi$_3$, 
\href{https://doi.org/10.1103/PhysRevB.100.014401}{Phys. Rev. B \textbf{100}, 014401 (2019).}

\bibitem{CePtSn2010MR} 
J. Prokle{\v{s}}ka, B. Detlefs, V. Sechovsk{\'y}, and M. M{\'i}{\v{s}}ek, 
Peculiarities of the magnetic-history-dependent phase in CePtSn, 
\href{https://doi.org/10.1016/j.jmmm.2009.09.017}{J. Magn. Magn. Mater. \textbf{322}, 1120 (2010).}

\bibitem{Dy3Co2000metamagnetic} 
N. V. Baranov, E. Bauer, R. Hauser, A. Galatanu, Y. Aoki, and H. Sato, 
Field-induced phase transitions and giant magnetoresistance in Dy$_3$Co single crystals, 
\href{https://doi.org/10.1007/s100510070250}{Eur. Phys. J. B \textbf{16}, 67 (2000).}

\bibitem{CePtSn2003neutron} 
B. Janou{\v{s}}ov{\'a}, V. Sechovsk{\'y}, K. Proke{\v{s}}, and T. Komatsubara, 
Microscopic origin of irreversible GMR effect in CePtSn around 3 T, 
\href{https://doi.org/10.1016/S0921-4526(02)01831-8}{Physica B \textbf{328}, 145 (2003).}

\end{thebibliography}

\end{document}